# *Photon momentum in surface plasmon*

V.S.Zuev and H.Ya.Zueva

*The P.N.Lebedev Physical Institute of RAS*
*53 Leninsky pr., 119991 Moscow, Russia*
**vizuev@sci.lebedev.ru**

A momentum of the photon in a surface plasmon could be hundreds and even thousand times higher as compared to a momentum of a photon in a free space. Such a photon could be scattered on an electron the later being supplied by a notable momentum. This momentum is approximately of the same value as a momentum transposed by an X-ray photon. A physical fundamental importance of the observation is outlined. Various practical applications of the phenomenon are proposed including an optical nanodevice.

# *Об импульсе фотона в поверхностном плазмоне*

В.С.Зуев, Г.Я.Зуева

Физический ин-т им. П.Н.Лебедева РАН
119991 Москва, Ленинский пр-т, 53
vizuev@sci.lebedev.ru

Импульс фотона в поверхностном плазмоне в зависимости от условий может в сотни и тысячу раз превышать импульс фотона на той же частоте в свободном пространстве. При рассеянии фотона из плазмона на электроне последний получает заметный импульс, сравнимый по величине с импульсом, передаваемым рентгеновским фотоном. Обращено внимание на физическую фундаментальность данного наблюдения. Предложены практические применения эффекта, в частности, в оптическом наноустройстве.

    Вопрос об импульсе электромагнитной волны в материальной среде обсуждается с ранних лет электромагнитной теории уже на протяжении столетия . Начало этому обсуждению, по-видимому, положили статьи Минковского /1/ и Абрагама /2/. В дискуссии участвовали многочисленные известные авторы – Дж.П.Гордон, В.Пайерлс, Д.В.Скобельцын и др. Современное состояние вопроса изложено в работе /3/, которая содержит множество ссылок по данному вопросу. Существо дела заключается в том, что тензор энергии-импульса электромагнитной волны не следует рассматривать без учета тензора энергии-импульса материальной среды. Коль скоро учет тензора энергии-импульса материальной среды проделан, то все рассмотрения оказываются, как правило, адекватными, а форма, в которой это проделывается, оказывается вопросом персонального выбора автора /3/.

    В рассмотрении, которое следует ниже, используются результаты работ /4,5/. В этих работах, как и в других работах, вопрос об импульсе электромагнитной волны рассмотрен применительно к пространству, заполненному однородным диэлектриком с диэлектрической проницаемостью $\varepsilon$. Импульс электромагнитной волны найден равным $\hbar k_\varepsilon$, где $k_\varepsilon$ - волновое число плоской электромагнитной волны в этом диэлектрике.

    Ниже будет рассмотрено иное однородное пространство – пространство с плоской границей раздела металл-диэлектрик. Как трехмерное это пространство неоднородно, однако как двумерное пространство оно является однородным. Более того, будут рассмотрены пространства с двумя и даже с четырьмя границами раздела металл-диэлектрик. Наличие дополнительных плоско-параллельных границ раздела не нарушает однородности двумерного пространства. В этих двумерных пространствах нет зависимости их свойств от продольных координат, и потому они оказываются однородными.

    Ниже будет также упоминаться пространство с тонкой нитью. Это пространство является одномерным однородным пространством.

    В этих пространствах существуют собственные волны (моды) электромагнитного поля, которых нет в 3-хмерном однородном пространстве. Ниже речь будет идти о волнах оптических частот, о волнах с энергией фотона $1-2$ эВ. Это будут поверхностные волны, называемые поляритон-плазмонами /6/. Плазмоны на тонких металлических пленках и нитях значительно замедлены по сравнению с волнами той

же частоты в свободном пространстве. Значение замедления может достигать $10^3$ /7,8/. Быть может существует и более значительное замедление, однако этот вопрос требует дополнительного исследования.

Анализ, подобный тому, что проделан в /4,5/, показывает, что в пространствах с пленкой и нитью импульс фотона в плазмоне равен $\hbar k_{pl}$, где $k_{pl} >> \omega/c$ - волновое число плазмона. Это означает, что оптический фотон в плазмоне с замедлением $10^3$ крат имеет импульс, равный импульсу рентгеновского фотона, фотона с энергией кванта $1-2$ кэВ. Рассеяние такого фотона на электроне будет сопровождаться последствиями, которые имеют место в эффекте Комптона и отсутствуют при томпсоновском рассеянии. Иными словами, рассеянный фотон должен отличаться сильно измененным импульсом, чего не происходит при томпсоновском рассеянии.

Эффект Комптона состоит в изменении длины волны фотона электромагнитного излучения вследствие рассеяния его электроном. Формулу Комптона часто пишут в следующем виде:

$$\lambda' - \lambda = \frac{h}{mc}(1 - \cos\theta), \qquad (1)$$

Здесь $\lambda$ - длина волны падающего фотона, $\lambda'$ - рассеянного фотона, $m$ - масса электрона, $c$ - скорость света в вакууме, $\theta$ - угол между направлениями движения падающего и рассеянного фотонов. Так называемая комптоновская длина волны $\lambda_c = h/mc$ для электрона равна $2.4263 \cdot 10^{-10}$ см. При рассеянии оптического фотона длины волны $500$ нм изменение длины волны составляет около $2.5 \cdot 10^{-3}$ нм, или $5 \cdot 10^{-4}$%.

Фотон в поверхностном плазмоне обладает до $10^3$ раз увеличенным импульсом. Поле поверхностной волны находится как внутри металла, так и вне его, оба – вблизи поверхности раздела, убывая экспоненциально при удалении от поверхности. При появлении электрона вблизи поверхности металла может произойти рассеяние фотона из плазмона на электроне. При рассеянии в волну свободного пространства изменение импульса, а, следовательно, энергии фотона мало. При рассеянии в плазмон происходит заметное изменение импульса и энергии рассеянного фотона. Расчет показывает, что энергия рассеянного фотона равна

$$\hbar\omega' \approx \hbar\omega[1 - n^2 \frac{\hbar\omega}{2mc^2}(1-\cos\theta)]. \qquad (2)$$

Первоначально электрон покоился. Импульсы падающего и рассеянного фотонов лежат в одной плоскости.

Получаемая из (2) пара значений $\hbar\omega'$ и $p_{\gamma'}$ в общем случае не соответствует резонансному плазмону на частоте $\omega'$. Нерезонансные плазмоны могут возбуждаться на границах раздела металл-диэлектрик, но при этом среди волн в пространстве с диэлектриком у границы должна присутствовать прилегающая волна, нарастающая при удалении от границы раздела /9/. На первый взгляд кажется, что это обстоятельство в произвольном случае запрещает возможность рассеяния в нерезонансный плазмон. Однако это не так. Следует предположить, что на удалении от рассматриваемой пленки металла находится вторая, параллельная пленка металла. Расстояние между пленками так велико, что можно не учитывать их взаимное влияние при расчете собственных мод полей. Однако с точки зрения возможного возбуждения плазмона во второй пленке наличие связи существенно. И хотя эта связь мала, ее малость проявляется лишь в малости величины сечения рассеяния фотона из плазмона на электроне, но не препятствует возбуждению плазмона во 2-й пленке. Заметим, что при выводе (2) кратность замедления на обоих частотах выбрана одинаковой.

Электрон приобретает кинетическую энергию, равную

$$(\vec{p}_{e'})^2 / 2m \approx n^2 \hbar\omega \frac{\hbar\omega}{2mc^2}(1-\cos\theta) \qquad (3)$$

Энергия покоя электрона $mc^2 \approx 0.51$ МэВ. При $\hbar\omega \approx 2$ эВ, $\theta = \pi$ и $n = 10^{2.55}$ получаем $(\vec{p}_{e'})^2 / 2m = 1$ эВ.

Дополнительного исследования требует вопрос об инерционности отклика среды.

Подведем итог. В сильно замедленной волне поверхностного плазмона фотон имеет импульс, увеличенный по сравнению с импульсом фотона в свободном пространстве в число раз, равное замедлению. Возможные значения замедления могут достигать $10^3$ раз. В такое же число раз увеличен импульс фотона.

При рассеянии фотона из плазмона на электроне последний получает значительный импульс и приобретает кинетическую энергию, значение которой может достигать половины энергии падающего фотона $\hbar\omega$. Рассеянный фотон имеет частоту $\omega'$, которая вдвое меньше частоты $\omega$ падающего фотона. Наблюдается эффект, аналогичный эффекту Комптона, то есть рассеяние света с изменением частоты, но в отличие от последнего – на оптической частоте.

Наблюдение рассеяния фотонов с изменением частоты (эффект Комптона) послужило в свое время доказательством наличия у электромагнитного излучения корпускулярных свойств. И тогда, и сейчас опыты проделывают с рентгеновскими и $\gamma$ - квантами. Для оптических квантов эффект мал - $0.1 \, см^{-1}$, хотя и доступен для измерения современными средствами. Конечно, нет сомнений в существование оптических фотонов. Однако опыты с фотонами из поверхностных плазмонов могут отчетливо продемонстрировать курпускулярные свойства света на оптических частотах.

Эти же опыты будут более однозначными в интерпретации, чем опыты в диэлектрических средах (см. обзор в /3/), имевшие целью выяснение вопроса об импульсе фотона в среде, но так и не давшие по данным из /3/ однозначного результата.

У рассмотренного явления могут быть и другие полезные применения. В наноустройствах с помощью электрона можно переключать излучение из одного наноустройства в другое. Может быть также построено устройство для генерирования пучка электронов с энергией $0.5 - 1 \, эВ$.

1. H.Minkowski. Nachr. Ges. Wiss. Goettingen, Math.-Phys., Kl, 53 (1908)
2. M.Abraham. Rend. Circ. Mat. Palermo, v.28, 1 (1909)
3. R.N.C.Pfeifer, T.A.Nieminen, N.R.Heckenberg, H.Rubinsztein-Dunlop. Rev. Mod. Phys., v.79, 1197-1216 (2007)
4. D.F.Nelson. Phys. Rev. A, v.44 3985-3996 (1991)
5. R.Loudon, L.Allen, D.F.Nelson. Phys. Rev. E, v.55, 1071-1085 (1997)
6. D.Pines. Rev. Mod. Phys., v.28, 184-198 (1956)
7. В.С.Зуев, Г.Я.Зуева. Оптика и спектроскопия, рег. № 28-08 от 15.02.08
8. V.S.Zuev, G.Ya.Zueva. J. Russian Laser Res., v.27, 167-184 (2006)
9. В.С.Зуев. Оптика и спектроскопия, т.102, 809-820 (2007)